\documentclass[12pt]{iopart}

\usepackage{graphicx,setspace}
\usepackage{dsfont}
\usepackage{iopams}
\usepackage{mathrsfs}
\usepackage{psfrag}
\usepackage{cases}

\begin{document}
\title{Damped vacuum states of light}

\author{T G Philbin}
\address{Physics and Astronomy Department, University of Exeter,
Stocker Road, Exeter EX4 4QL, UK.}
\eads{\mailto{t.g.philbin@exeter.ac.uk}}

\begin{abstract}
We consider one-dimensional propagation of quantum light in the presence of a block of material, with a full account of dispersion and absorption. The electromagnetic zero-point energy for some frequencies is damped (suppressed) by the block below the free-space value, while for other frequencies it is increased. We also calculate the regularized (Casimir) zero-point energy at each frequency and find that it too is damped below the free-space value (zero) for some frequencies. The total Casimir energy is positive.
\end{abstract}

\section{Introduction}
The most fascinating aspect of the Casimir effect is that it attributes a physical significance to the vacuum zero-point energy of electromagnetic radiation~\cite{milonni}. This is all the more surprising as electromagnetic zero-point energy seems initially to be a theoretical malfunction, a spurious divergence that should be ignored. An infinite zero-point energy for quantum fields is a simple consequence of their description as a continuum of quantum harmonic oscillators, where all oscillation frequencies may occur~\cite{milonni,loudon}. Casimir theory attributes a real existence to a finite part of the electromagnetic zero-point energy and pressure, a part that is determined by macroscopic objects and that in turn exerts forces on those objects~\cite{lif55,dzy61,LLsp2,dalvit,simpson}. Experimental confirmation of Casimir forces allows us to take seriously an electromagnetic vacuum state whose energy density and pressure have a spatial and frequency structure that is determined by the electromagnetic susceptibilities of macroscopic materials. But the detailed structure of the electromagnetic vacuum state is not usually described in Casimir calculations because only the total vacuum energy, or total vacuum pressure at material boundaries, is required to find the forces. Moreover the sum over all frequency contributions to the force is invariably carried out as a sum over imaginary frequencies, which obscures the spectrum of the vacuum state since each imaginary-frequency component has contributions from all real frequencies. Here we study the electromagnetic vacuum state in the simplest non-trivial case and show that individual modes of light can have zero-point energies that are damped (decreased) by coupling to macroscopic materials. These damped vacuum states of light are an interesting addition to the single-mode states routinely discussed in the quantum-optics textbooks~\cite{loudon}. 
 
Most treatments of Casimir energy are focussed on the total energy rather than on the energy of individual modes~\cite{mil14,gra03,kli15}. Moreover idealized boundary conditions are often assumed~\cite{mil14,gra03}. In~\cite{for88,ell08} the Casimir energy and stress of individual frequencies was studied but only by making use of idealized material properties that violate Kramers-Kronig relations. Here we consider dielectric functions that exhibit dispersion and absorption consistent with Kramers-Kronig relations, as is required for accurate theoretical predictions of the Casimir effect.

Casimir energy quantifies forces between objects through its derivative with respect to the separation distances of the objects. The Casimir force is also determined by the zero-point electromagnetic stress tensor, which is the most common method for computing the force~\cite{lif55,dzy61,LLsp2}. It is important to note that the Casimir energy can be calculated separately from the stress tensor, by means of the conserved quantity associated with time-translation symmetry (for non-moving materials)~\cite{phi11}. The Casimir energy obtained in this manner gives a force that agrees with that obtained from the stress tensor~\cite{phi11}, but part of the energy does not contribute to the force. This last fact is due to a self-energy contribution from material inhomogeneities (including sharp boundaries) that does not cause a force between separated objects. A complete description of Casimir energy must include such self-energy contributions, as they are necessarily present if one adopts the standard view that time-translation symmetry gives energy through Noether's theorem. The role of self-energy contributions has been recognised in discussions of the gravitational effects of Casimir energy~\cite{mil14}. Here we will consider the zero-point energy of individual (real) frequencies, including self-energy contributions. A regularization of the zero-point energy is required to obtain the physical Casimir energy as the total zero-point energy always diverges. As we consider realistic materials that obey Kramers-Kronig relations, the regularizaton is the standard one employed in the prediction of Casmir forces through the stress tensor~\cite{lif55,dzy61,LLsp2}. 
 
Casimir forces are intimately connected to thermal radiation~\cite{lif55,dzy61,LLsp2} and our approach here is similar to investigations of the effect of material boundaries on the thermal spectrum~\cite{wil06,jou05}. Also closely related are studies of the spatial variation in the local electromagnetic density of states caused by materials, an effect that can be probed by spontaneous emission~\cite{pur46,dre68,bar98,Novotny}. In those investigations, it is the properties of light outside the materials, including close to the boundaries, that are relevant. Our interest here however is in the energy density of each mode throughout space, which crucially includes the regions inside the materials. 

We are also motivated by the developing subject of nanomechanical systems~\cite{oco10,teu11,cha11,poo12,gro13}. The physics of such systems may be modelled as quantum oscillators that are damped by coupling to reservoirs representing the environment. Although the link to nanomechanical systems is not often made, the quantum theory of macroscopic electromagnetism is also a theory of quantum damped harmonic oscillators and can be formulated exactly as a quantized theory of light coupled to a reservoir~\cite{bha06,khe06,sut07,amo08,khe10,phi10}. Results for the behavior of quantum light modes in macroscopic electromagnetism are therefore instructive for studies of nanomechanical systems.

\begin{figure}[!htbp]
\begin{center} 
\includegraphics[width=9cm]{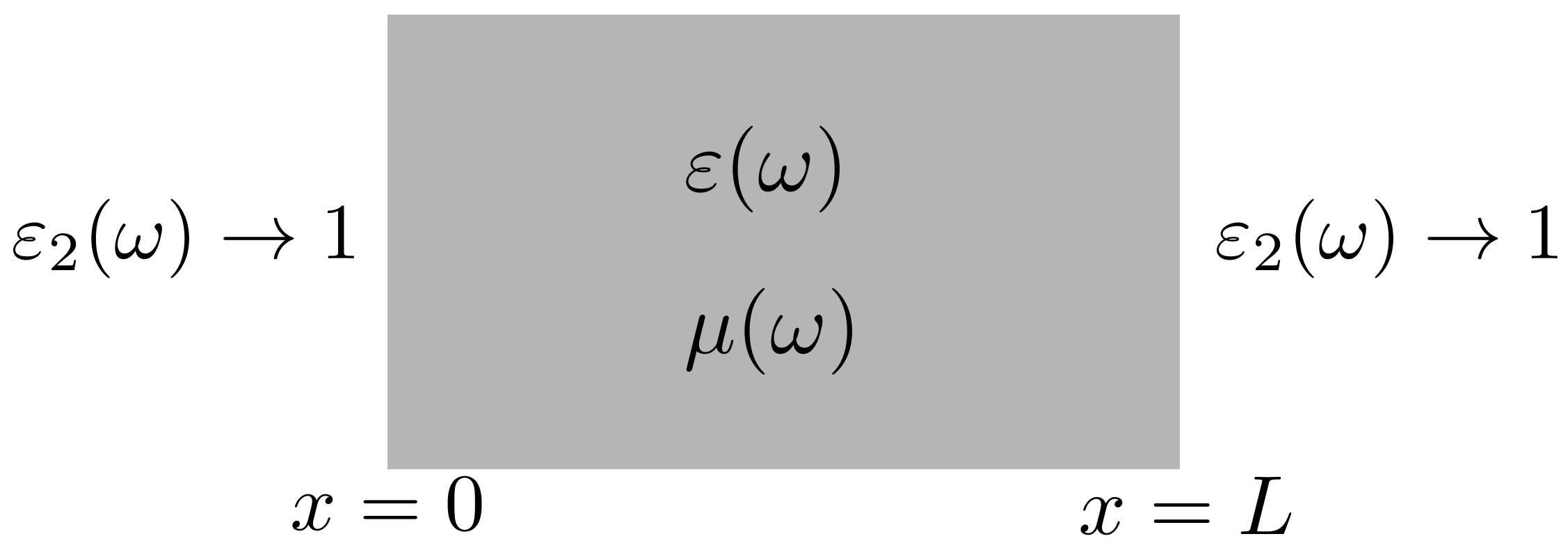}
\caption{A block of material with electric permittivity $\varepsilon(\omega)$ and magnetic permeability $\mu(\omega)$. The block extends infinitely in the $y$ and $z$ directions and has boundaries at $x=0$ and $x=L$. The surrounding region is vacuum, written as the vacuum limit $\varepsilon_2(\omega)\to 1$ of another material. We consider light linearly polarized in the $y$-direction and propagating in the $x$-direction perpendicular to the block faces at $x=0$ and $x=L$.}
\label{fig:block}
\end{center}
\end{figure}

\section{Set-up}
We consider a block with electric permittivity $\varepsilon(\omega)$ and magnetic permeability $\mu(\omega)$ (see Fig.~\ref{fig:block}). The block has boundaries at $x=0$ and $x=L$ and is surrounded by vacuum, but for technical reasons (see below) the vacuum region is represented as the vacuum limit $\varepsilon_2(\omega)\to 1$ of a different dielectric. Our interest is in the effect of this block on the quantum vacuum state of light, and to simplify the analysis as much as possible we consider only light modes linearly polarized in the $y$-direction and propagating in the $x$-direction. The set-up is therefore essentially one-dimensional and we treat it as such. The input and output modes in this arrangement are analysed in~\cite{gru96}.

The macroscopic Maxwell equations for light interacting with an arbitrary inhomogeneous material whose dielectric functions obey the Kramers-Kronig relations can be formulated exactly as closed system of electromagnetic fields coupled to a reservoir (see~\cite{phi10}, for example). By quantizing this system and diagonalizing its Hamiltonian~\cite{phi10} one obtains the following expression for the electric-field operator in our case of interest, where the general result is specialized to one polarization and to the case where both propagation and material inhomogeneity are in the $x$-direction:
\begin{eqnarray}
\fl
\hat{E}(x,t)=\sqrt{\frac{\hbar\mu_0}{\pi}}\int_0^\infty \rmd\omega\int_{-\infty}^\infty \rmd x' &\, g(x,x',\omega)  \left[-\frac{\omega^2}{c}\sqrt{ \varepsilon_\mathrm{I}(x',\omega)} \hat{C}_\mathrm{e}(x',\omega) \right.  \nonumber  \\
&  \left. +\rmi \omega \partial_{x'}\left( \sqrt{- \kappa_\mathrm{I}(x',\omega)} \hat{C}_\mathrm{m}(x',\omega) \right) \right] e^{-\rmi \omega t} +\mathrm{h.c.}   \label{Eop}
\end{eqnarray}
Here the notation is as follows: $\varepsilon_\mathrm{I}(x,\omega)$ is the imaginary part of the inhomogeneous permittivity $\varepsilon(x,\omega)$,  $\kappa_\mathrm{I}(x,\omega)$ is the imaginary part of $1/\mu(x,\omega)$, h.c.\ means hermitian conjugate,  $\hat{C}_\mathrm{e}(x,\omega)$ and $\hat{C}_\mathrm{m}(x,\omega)$ are the annihilation operators for the normal modes that diagonalize the Hamiltonian and obey
\begin{equation}
\left[ \hat{C}_\mathrm{\lambda}(x,\omega),\hat{C}^\dagger_\mathrm{\lambda'}(x',\omega') \right] =\delta_{ \lambda\lambda'} \delta(x-x')  \delta(\omega-\omega'), \qquad \lambda,\lambda'=\{\mathrm{e},\mathrm{m}\},
\end{equation}
and $g(x,x',\omega)$ is the retarded Green function satisfying
\begin{equation}  \label{geqn}
\left(\partial_{x}\frac{1}{\mu(x,\omega)}\partial_{x}+k_0^2\varepsilon(x,\omega)\right)g(x,x',\omega)= \delta(x-x') , \qquad k_0=\frac{\omega}{c} .
\end{equation}
In the arrangement of Fig.~\ref{fig:block} the material is piecewise homogeneous and the solution of (\ref{geqn}) is well known (see~\cite{gru96} for example). The Green function determines the electric-field operator (\ref{Eop}) and the magnetic-field operator (which points in the $z$-direction) is found from $\hat{B}(x,\omega)=-\rmi \partial_{x}\hat{E}(x,\omega)/\omega$. Because of the imaginary parts of the dielectric functions in (\ref{Eop}), it is necessary to consider vacuum as the limit of a  permittivity going to $1$; this limit is to be taken in final observable quantities, or earlier if this will not affect the final results.

In the absence of materials, the field operators for one-dimensional propagation can be decomposed into independent left- and right-going modes of each frequency. In our case, reflections from the block boundaries couple the left- and right-going modes. Nevertheless, in the vacuum regions outside the block the field operators and their algebra still take a relatively simple form (see also~\cite{gru96}).  In the region to the right of  the block ($x>L$) the electric-field operator (\ref{Eop}) can be written
\begin{equation}
\fl
\hat{E}(x,t)=\int_0^\infty \rmd\omega \sqrt{\frac{\hbar\omega}{4\pi c\varepsilon_0}}  \left[  e^{\rmi \sqrt{\varepsilon_2(\omega)}\, k_0 x} \hat{a}_+(x,\omega)  + e^{-\rmi \sqrt{\varepsilon_2(\omega)}\, k_0 x} \hat{a}_-(x,\omega) \right] e^{-\rmi \omega t} +\mathrm{h.c.}  \label{Eopr}   
\end{equation}
The operators for the right-going ($+$) and left-going ($-$) modes in (\ref{Eopr}) have the following simple algebra in the vacuum limit $\varepsilon_2(\omega)\to 1$ outside the block, where we denote this limit by an arrow:
\begin{eqnarray}
\fl
\left[ \hat{a}_+(x,\omega) , \hat{a}^\dagger_+(x,\omega) \right] \to \delta(\omega-\omega'), \qquad  \left[ \hat{a}_-(x,\omega) , \hat{a}^\dagger_-(x,\omega) \right] \to \delta(\omega-\omega'),  \label{aal1}  \\
\fl
\left[ \hat{a}_+(x,\omega) , \hat{a}_-(x,\omega) \right] \to 0, \qquad  \left[ \hat{a}_+(x,\omega) , \hat{a}^\dagger_-(x,\omega) \right] \to  \rmi \zeta(\omega,L)  \delta(\omega-\omega'),    \label{aal2}   \\
 \zeta(\omega,L) = \frac{\rmi e^{-2 i k_0 L } (\varepsilon -\mu )}{ 2 i n  \cot \left( n
    k_0L \right) +\varepsilon+\mu}.   \label{gamma}
\end{eqnarray}
Here $n=\sqrt{\varepsilon\mu}$ is the refractive index of the block and we have suppressed the frequency dependence of $\varepsilon$, $\mu$ and $n$. The failure of all the right ($+$) operators to commute with all the left ($-$) operators in (\ref{aal1}) and (\ref{aal2}) shows the coupling of these modes due to reflection from the block boundaries. Independent modes with commuting operators can be defined as follows, where $\hat{b}_1$ and $\hat{b}_2$ are the associated annihilation operators:
\begin{eqnarray}
 \hat{b}_1(x,\omega)=\frac{1}{2}\left[ \left(   \delta_+ + \delta_-  \right)  e^{-\rmi\phi_\zeta/2}  \hat{a}_+(x,\omega) + \rmi  \left(   \delta_+ - \delta_-  \right) e^{\rmi\phi_\zeta/2}  \hat{a}_-(x,\omega) \right] ,  \\
 \hat{b}_2(x,\omega)=\frac{1}{2}\left[ \left(   \delta_+ + \delta_-  \right)  e^{\rmi\phi_\zeta/2}  \hat{a}_-(x,\omega) - \rmi  \left(   \delta_+ - \delta_-  \right) e^{-\rmi\phi_\zeta/2}  \hat{a}_+(x,\omega) \right] ,   \\
\delta_+=\left(1+|\zeta|\right)^{-1/2}, \qquad \delta_-=\left(1-|\zeta|\right)^{-1/2}, 
\end{eqnarray}
where the function (\ref{gamma}) is decomposed as $\zeta=|\zeta|e^{\rmi \phi_\zeta}$. Inside the block the algebra of field operators is more complicated but is straightforwardly obtained from (\ref{Eop}) and the Green function $g(x,x',\omega)$.

\section{Vacuum-state field uncertainties}
The electromagnetic vacuum state is defined by $\hat{C}_\mathrm{e}(x,\omega)|0\rangle=0$ and $\hat{C}_\mathrm{m}(x,\omega)|0\rangle=0$, which imply $\hat{a}_\pm(x,\omega)|0\rangle=0$ for the mode operators in (\ref{Eopr}). It is then straightforward to calculate the zero-point uncertainties of the electric ($\Delta E(x)$) and magnetic ($\Delta B(x)$) fields in the region to the right of  the block ($x>L$) using the algebra (\ref{aal1}) and (\ref{aal2}):
\begin{eqnarray}
\left[\Delta E(x)\right]^2=\langle0|[\hat{E}(x,t)]^2|0\rangle=\frac{\hbar c\mu_0}{2\pi} \int_0^\infty \rmd\omega \, \omega\left[1-|\zeta| \sin\left(2k_0x+\phi_\zeta\right)\right],  \label{DEvac} \\
\left[\Delta B(x)\right]^2=\langle0|[\hat{B}(x,t)]^2|0\rangle=\frac{\hbar \mu_0}{2\pi c} \int_0^\infty \rmd\omega \, \omega\left[1+|\zeta| \sin\left(2k_0x+\phi_\zeta\right)\right].  \label{DBvac}
\end{eqnarray}
These field uncertainties oscillate with distance from the block boundary. The electric-field uncertainty (\ref{DEvac}) can be probed by measuring spontaneous emission at different positions (though in general modes for all angles of incidence on the block need to be included)~\cite{pur46,dre68,bar98,Novotny}.

The vacuum-state field uncertainties $\Delta E(x)$ and $\Delta B(x)$ for a general inhomogeneous block have a simple expression in terms of the Green function:
\begin{eqnarray}
\left[\Delta E(x)\right]^2=\langle0|[\hat{E}(x,t)]^2|0\rangle=-\frac{\hbar \mu_0}{\pi} \mathrm{Im} \int_0^\infty \rmd\omega \, \omega^2 g(x,x,\omega) ,  \label{DEgen} \\
\left[\Delta B(x)\right]^2=\langle0|[\hat{B}(x,t)]^2|0\rangle=-\frac{\hbar \mu_0}{\pi} \mathrm{Im} \int_0^\infty \rmd\omega \,  \lim_{x' \to x} \partial_{x}\partial_{x'} g(x,x',\omega),  \label{DBgen}
\end{eqnarray}
which follow from (\ref{Eop}) and (\ref{geqn}) (see~\cite{phi11} for example). These expressions are exactly what would be expected as the extension to the quantum vacuum of the classical results of Rytov for the field variances of thermal radiation~\cite{rytov,lif55,LLsp2}. The results (\ref{DEvac}) and (\ref{DBvac}) can also be obtained from the general expressions (\ref{DEgen}) and (\ref{DBgen}), as the Green function to the right of the block ($x>L$, $x'>L$) with $\varepsilon_2(\omega)\to 1$ is
\begin{equation}
g(x,x',\omega)=-\frac{\rmi}{2 k_0}e^{\rmi k_0|x-x'|}+\frac{1}{2k_0}\zeta e^{\rmi k_0(x+x')},
\end{equation}
where $\zeta$ is again (\ref{gamma}). 

To obtain the field uncertainties inside the block we can use (\ref{DEgen}) and (\ref{DBgen}) together with the Green function in that region ($0<x<L$, $0<x'<L$). The Green function here is ($\varepsilon_2(\omega)\to 1$):
\begin{eqnarray}
\fl
g(x,x',\omega)= -\frac{\rmi\mu}{2n k_0} \left[  e^{\rmi n k_0|x-x'|}+\alpha(\omega,L)\left( e^{\rmi nk_0(x-x')} +e^{-\rmi nk_0(x-x')}\right) \right.  \nonumber \\
\qquad\quad  \left.+\beta(\omega,L) \left( e^{\rmi nk_0(x+x')} + e^{-\rmi nk_0(x+x'-2L)}\right) \right],  \\
\alpha(\omega,L)=\left[ \left(\frac{n+\mu}{n-\mu}\right)^2 e^{-2\rmi n k_0 L} -1\right]^{-1} ,  \label{alpha}  \\\beta(\omega,L)=\frac{\varepsilon-\mu}{2n+\varepsilon+\mu+(2n-\varepsilon-\mu)e^{2\rmi n k_0 L} }, \label{beta}
\end{eqnarray}
giving the following field uncertainties inside the block:
\begin{eqnarray}
\left[\Delta E(x)\right]^2=\frac{\hbar c \mu_0}{2\pi} \mathrm{Im} \int_0^\infty \rmd\omega \, \frac{\rmi\mu \omega }{n}\left[ 1+2\alpha +\beta\left(e^{2\rmi nk_0x} +e^{-2\rmi nk_0(x-L)} \right) \right],  \label{DEmat} \\
\left[\Delta B(x)\right]^2=\frac{\hbar \mu_0}{2\pi c} \mathrm{Im} \int_0^\infty \rmd\omega \, \rmi\mu n \omega \left[ 1+2\alpha -\beta\left(e^{2\rmi nk_0x} +e^{-2\rmi nk_0(x-L)} \right) \right].  \label{DBmat}
\end{eqnarray}

\section{Vacuum-state energy}  \label{sec:energy}
When macroscopic electromagnetism is formulated as a closed system of electromagnetic fields coupled to  a reservoir, the total energy-density operator follows from time-translation symmetry and Noether's theorem~\cite{phi11}. The total energy has ``free" electromagnetic and reservoir terms, and also terms involving the coupling functions~\cite{phi11}.  For the ground state and thermal equilibrium, the electromagnetic part of the energy can be defined using the Hamiltonian of mean force~\cite{kir35,jar04,Campisi09,sub12,HSAL11,phi15}. The resulting electromagnetic energy is the total energy minus the energy the reservoir would have if it alone were present (for the vacuum state this is also the expectation value of the Hamiltonian of mean force). This definition of energy gives the same answer for Casimir forces as obtained by using the electromagnetic stress tensor~\cite{phi11}. In our case of one polarization and propagation in the $x$-direction, the electromagnetic energy per unit length $\rho(x)$ of the vacuum state in an arbitrary inhomogeneous material is~\cite{phi11}
\begin{equation}
\fl
\rho(x)=\frac{\varepsilon_0}{2} \mathrm{Im} \int_0^\infty \rmd\omega \left\{ \frac{\rmd[\omega\varepsilon(x,\omega)]}{\rmd \omega}\left[\Delta E(x)\right]^2 +\frac{c^2}{[\mu(x,\omega)]^2} \frac{\rmd[\omega\mu(x,\omega)]}{\rmd \omega}\left[\Delta B(x)\right]^2 \right\},  \label{rhogen}
\end{equation}
where $\Delta E(x)$ and $\Delta B(x)$ are the vacuum-state field uncertainties given by (\ref{DEgen}) and (\ref{DBgen}).

Using (\ref{rhogen}), we obtain the zero-point electromagnetic energy per unit length in the presence of the block. In the region to the right of the block ($x>L$), (\ref{DEvac}) and (\ref{DBvac}) together with $\varepsilon(x,\omega)=\mu(x,\omega)=1$ in (\ref{rhogen}) give
\begin{equation} \label{rhoempty}
\rho(x)=\frac{\hbar}{2\pi c} \mathrm{Im} \int_0^\infty \rmd\omega \,\rmi \omega,
\end{equation}
which is exactly the (diverging) zero-point energy per unit length in the absence of the block. This is also the result in the region to the left of the block ($x<0$). Thus, although the electric and magnetic field uncertainties outside the block are affected by its presence, this material dependence cancels out in the energy per unit length outside the block. It should be noted that this cancelation only happens for modes propagating perpendicular to the block boundaries. As our interest is in how the block alters the electromagnetic zero-point energy (of $x$-propagating modes), we see that only the energy inside the block matters. Inserting  (\ref{DEmat}) and (\ref{DBmat}) in (\ref{rhogen}) and putting $\varepsilon(x,\omega)=\varepsilon(\omega),$ $\mu(x,\omega)=\mu(\omega)$, we find the energy per unit length inside the block ($0<x<L$). Integration from $x=0$ to $x=L$ then gives the zero-point energy $\mathcal{E}$ in the block, and the result is
\begin{eqnarray}
 \mathcal{E}  =  \int_0^\infty \rmd\omega \, W(\omega),     \label{entot} \\
W(\omega)=\frac{\hbar \omega}{2\pi c} \mathrm{Im}   \left[ \rmi L(1+2\alpha)\frac{\rmd (\omega n)}{\rmd \omega}  
  +c \beta\left(e^{2\rmi nk_0L}-1\right)\frac{\mu}{n}  \frac{\rmd}{\rmd \omega} \left(\frac{n}{\mu}   \right)  \right], \label{W}
\end{eqnarray}
where $\alpha$ and $\beta$ are again (\ref{alpha}) and (\ref{beta}) and we have defined the spectral energy $W(\omega)$, i.e. the energy per unit frequency. The zero-point energy (\ref{entot}) of $x$-propagating modes contained in the block diverges because of the integral over mode frequencies, whereas the spectral energy (\ref{W}) is finite for each frequency. (If we integrated over all angles of incidence to the boundaries then the spectral energy would diverge; this divergence is due to the fact that spatial dispersion is not included in our dielectric functions~\cite{hor14}.) 

For the modes considered here our results show that, for each frequency, the block causes a finite change in the zero-point energy. The only change in the zero-point energy occurs inside the block, where modes with a small frequency spread $\Delta\omega$ around $\omega$ have a zero-point energy $\Delta\omega W(\omega)$, whereas without the block their energy in this region would be $\Delta\omega\hbar \omega L/(2\pi c)$ (the spectral energy per unit length of empty space is, from (\ref{rhoempty}), $\hbar \omega /(2\pi c)$). Thus $W(\omega)-\hbar \omega L/(2\pi c)$ quantifies the (finite) change in zero-point energy for each frequency. This finite result at each frequency is obtained without having to regularize any diverging quantities. Because the total zero-point energy diverges (even for the limited set of modes we consider), regularization is needed to compute finite Casimir energies and forces (see below). Regularization changes the zero-point energy attributed to each mode, but the change in zero-point energy caused by the block is finite for each mode both before and after regularization.

The zero-point energy (\ref{entot}) contains a diverging part $\mathcal{E}_\mathrm{bulk}$ that  is independent of the block boundaries, i.e. it is $L$ times the energy per unit length in an infinite material of refractive index $n$:
\begin{equation}  \label{enbul}
\mathcal{E}_\mathrm{bulk}=\frac{\hbar L}{2\pi c} \mathrm{Im} \int_0^\infty \rmd\omega \, \rmi\omega  \frac{\rmd (\omega n)}{\rmd \omega}  .
\end{equation}
(The zero-point energy per unit length in an infinite homogeneous material differs from (\ref{rhoempty}) by having a factor $\rmd (\omega n)/\rmd \omega$ in the integrand, as follows from (\ref{rhogen}).) In Casimir theory, all such diverging bulk quantities are dropped, and only the finite quantities that remain are taken as physically significant~\cite{lif55,dzy61,LLsp2}. The rationale for this regularization procedure is that only macroscopic material inhomogeneities (smooth or discontinuous changes in the dielectric functions) give physical meaningful contributions to electromagnetic zero-point quantities~\cite{LLsp2}. A notable feature of the regularization is that \emph{different} infinite quantities are dropped at different points of space (the diverging bulk contribution at any point depends on the values of the dielectric functions at that point)~\cite{LLsp2}. Thus, to obtain the Casimir energy $\mathcal{E}_C$ of the modes considered here, in the presence of the block, we drop the purely bulk quantity (\ref{rhoempty}) outside the block and remove (\ref{enbul}) from (\ref{entot}), i.e. $\mathcal{E}_C=\mathcal{E}-\mathcal{E}_\mathrm{bulk}$, so that
 \begin{eqnarray}
 \mathcal{E}_C  =  \int_0^\infty \rmd\omega \, W_C(\omega),  \label{encas} \\
W_C(\omega)=\frac{\hbar \omega}{2\pi c} \mathrm{Im}   \left[ 2\rmi L\alpha \frac{\rmd (\omega n)}{\rmd \omega}  
  +c \beta\left(e^{2\rmi nk_0L}-1\right)\frac{\mu}{n}  \frac{\rmd}{\rmd \omega} \left(\frac{n}{\mu}   \right)  \right],         \label{WC}
\end{eqnarray}
where we have defined the Casimir spectral energy $W_C(\omega)$, the Casimir energy per unit frequency.  As noted in the Introduction, the regularization employed in obtaining (\ref{WC}) is the standard one used to predict experimentally measured Casimir forces~\cite{lif55,dzy61,LLsp2}.

\begin{figure}[!htbp]
\begin{center} 
\includegraphics[width=9cm]{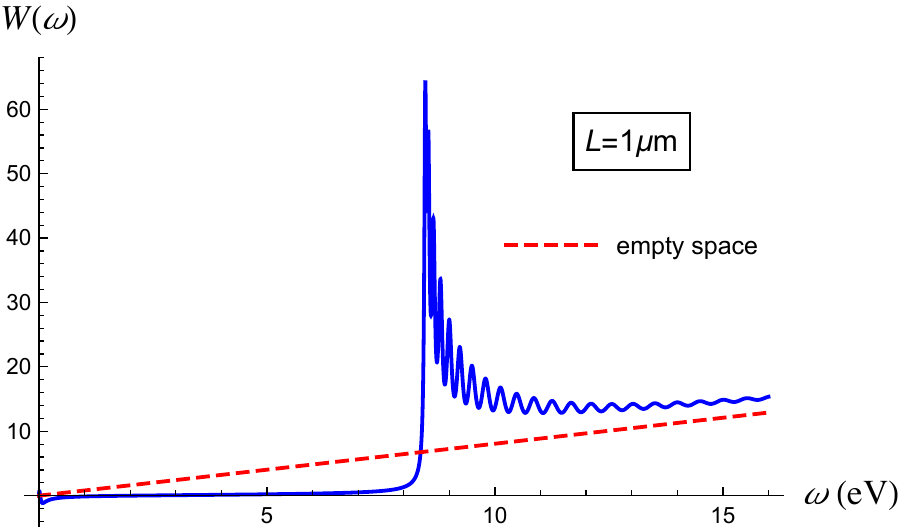}

\vspace{2mm}

\includegraphics[width=9cm]{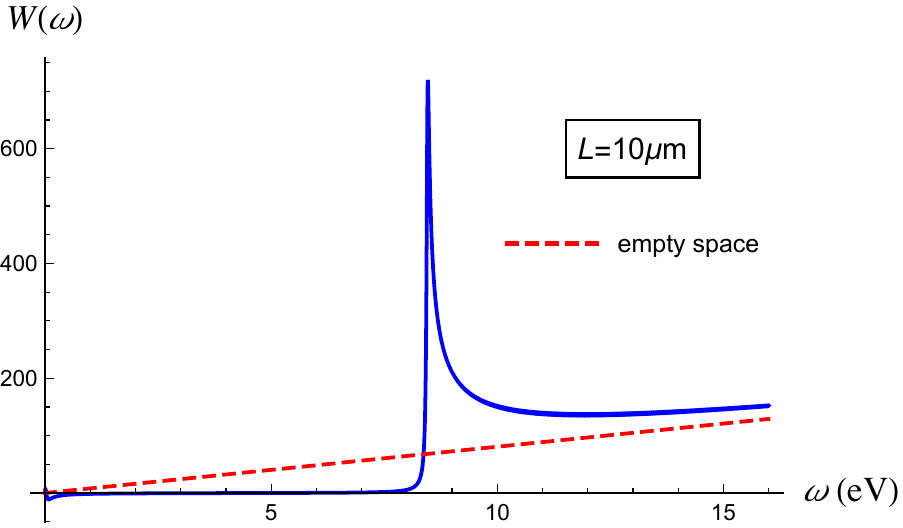}
\caption{Electromagnetic zero-point energy per unit frequency (\ref{W}) inside a metal of length $L$ (blue curves). The material has permittivity (\ref{osc}) with $\omega_0=0$, $\Omega=8.45\,\mathrm{eV}$ and $\gamma=0.047\,\mathrm{eV}$. The dashed red lines are $\hbar \omega L/(2\pi c)$, the value of $W(\omega)$ in the same spatial region  but without the block. In the top plot $L=1\,\mu\mathrm{m}$, in the bottom plot $L=10\,\mu\mathrm{m}$. The zero-point energy is less than the free-space value for frequencies $\omega\lesssim\Omega$.}
\label{fig:gold1}
\end{center}
\end{figure}

\begin{figure}[!htbp]
\begin{center} 
\includegraphics[width=9cm]{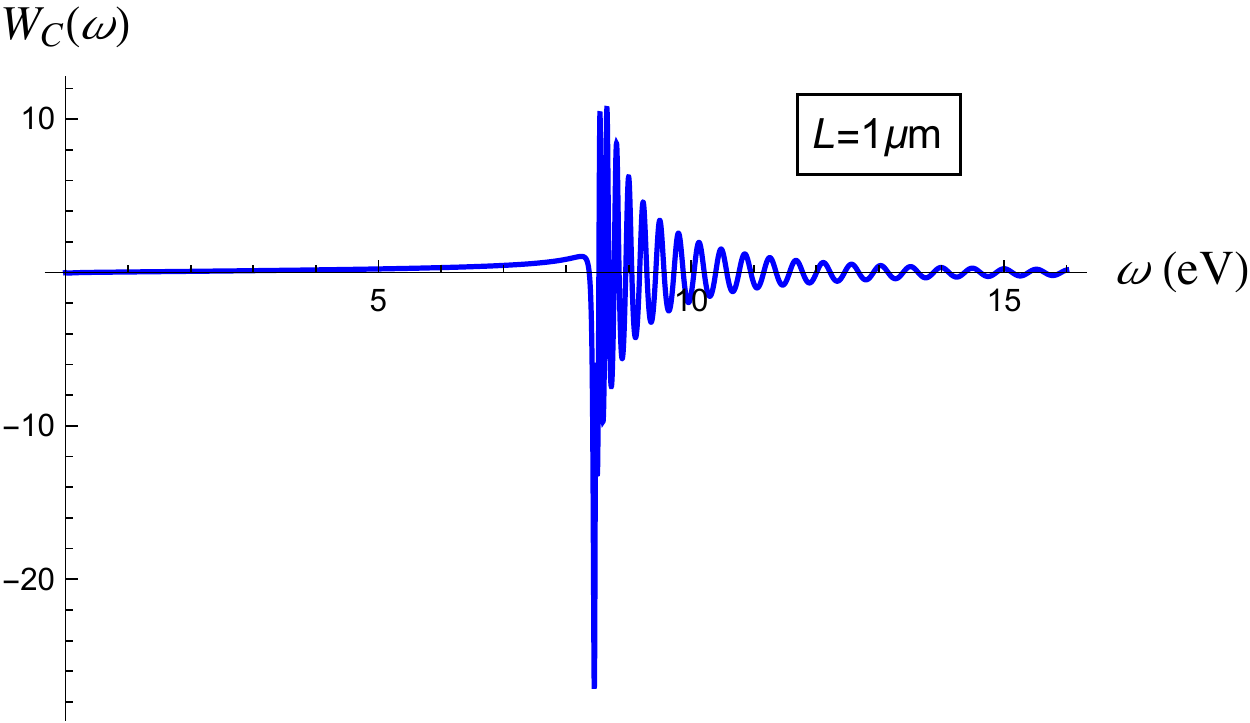}

\vspace{2mm}

\includegraphics[width=9cm]{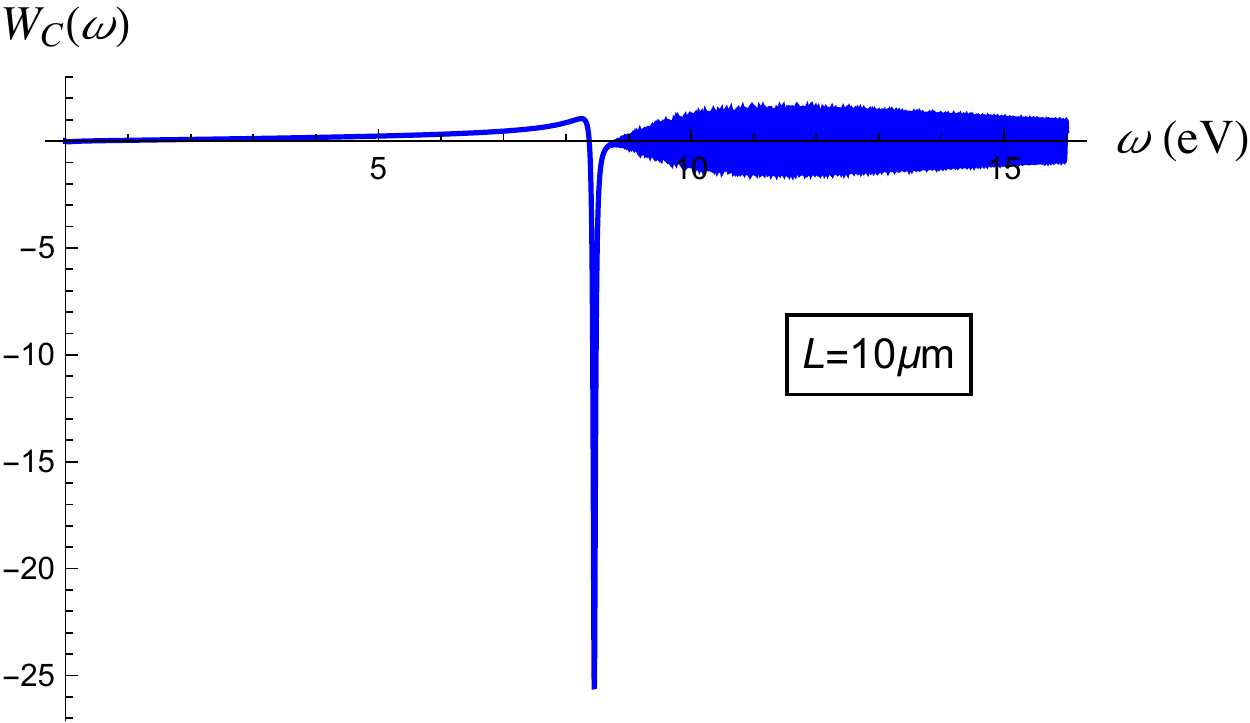}
\caption{Casimir energy per unit frequency (\ref{WC}) of the metal blocks described in Fig.~\ref{fig:gold1}. The total Casimir energy (\ref{encas}) is positive for all lengths $L$.}
\label{fig:gold2}
\end{center}
\end{figure}

\begin{figure}[!htbp]
\begin{center} 
\includegraphics[width=9cm]{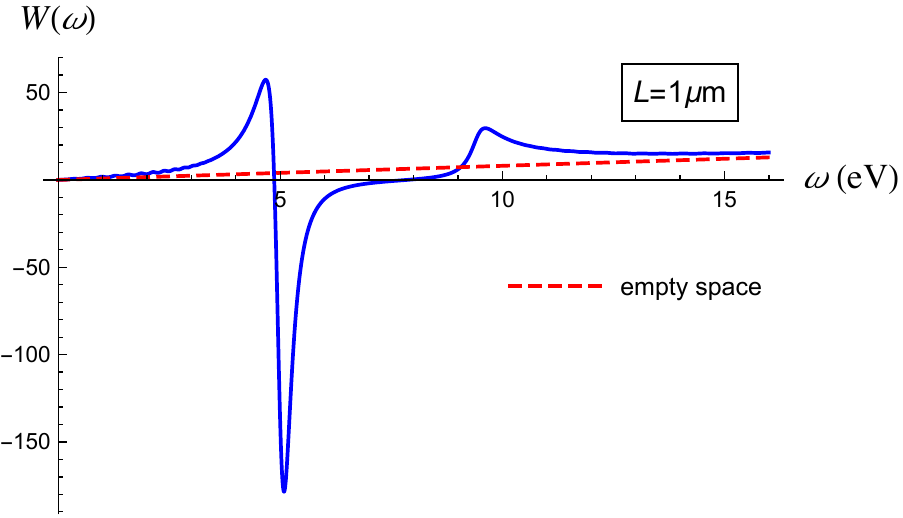}

\vspace{2mm}

\includegraphics[width=9cm]{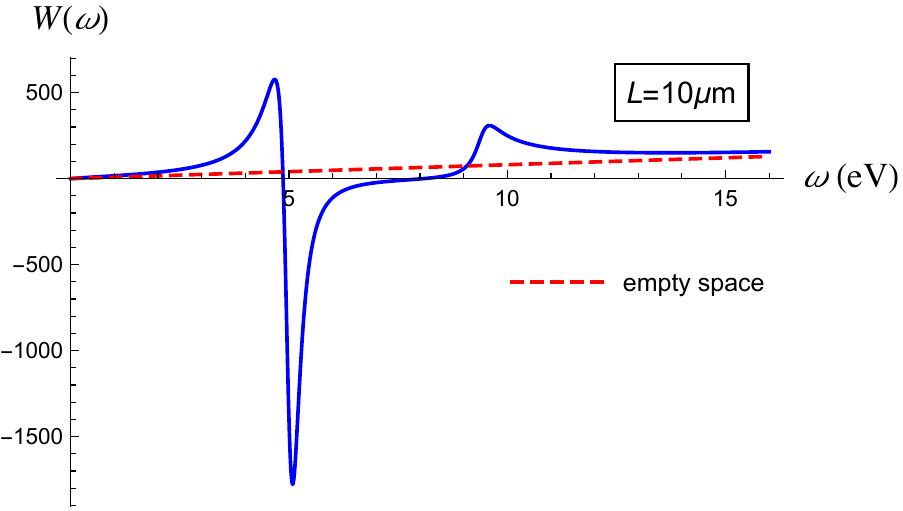}
\caption{Electromagnetic zero-point energy per unit frequency (\ref{W}) inside a non-metallic dielectric (blue curves) of length $L$. The material has permittivity (\ref{osc}) with $\omega_0=5\,\mathrm{eV}$, $\Omega=8\,\mathrm{eV}$ and $\gamma=0.5\,\mathrm{eV}$.  The dashed red lines are $\hbar \omega L/(2\pi c)$, the value of $W(\omega)$ in the same spatial region but with the block replaced by empty space. In the top plot $L=1\,\mu\mathrm{m}$, in the bottom plot $L=10\,\mu\mathrm{m}$.}
\label{fig:diel1}
\end{center}
\end{figure}

\begin{figure}[!htbp]
\begin{center} 
\includegraphics[width=9cm]{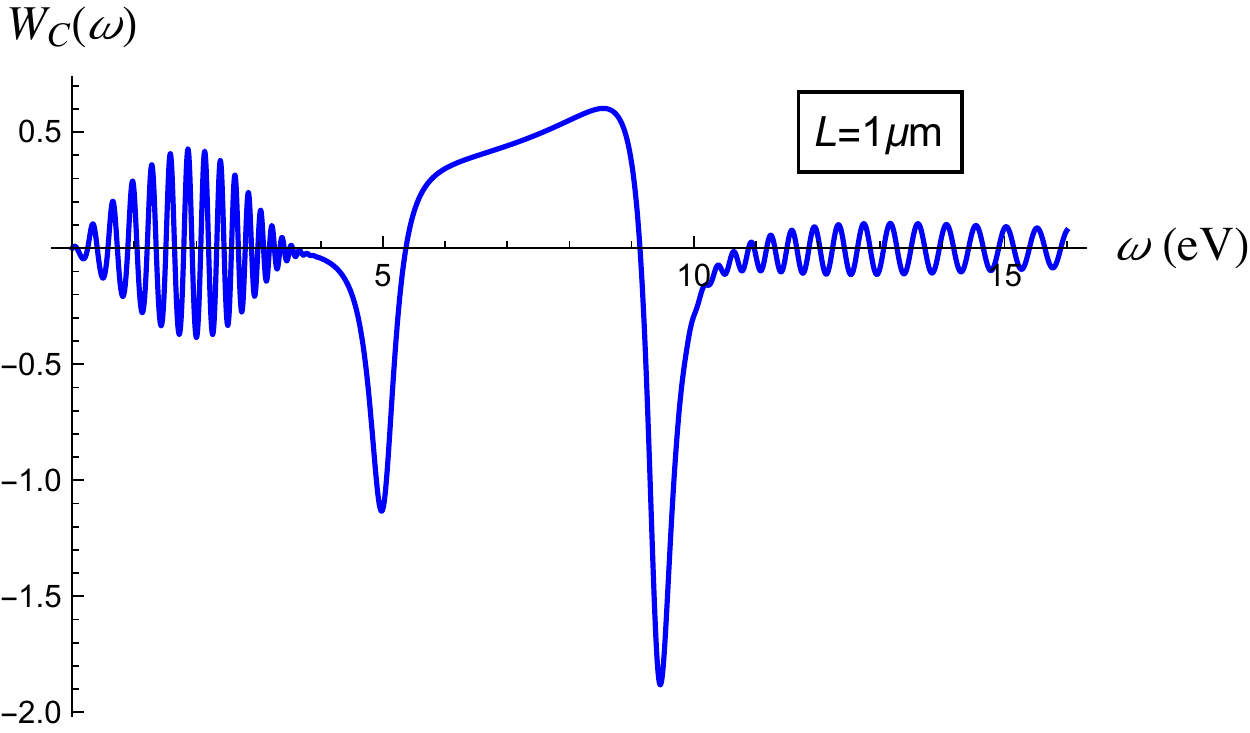}

\vspace{2mm}

\includegraphics[width=9cm]{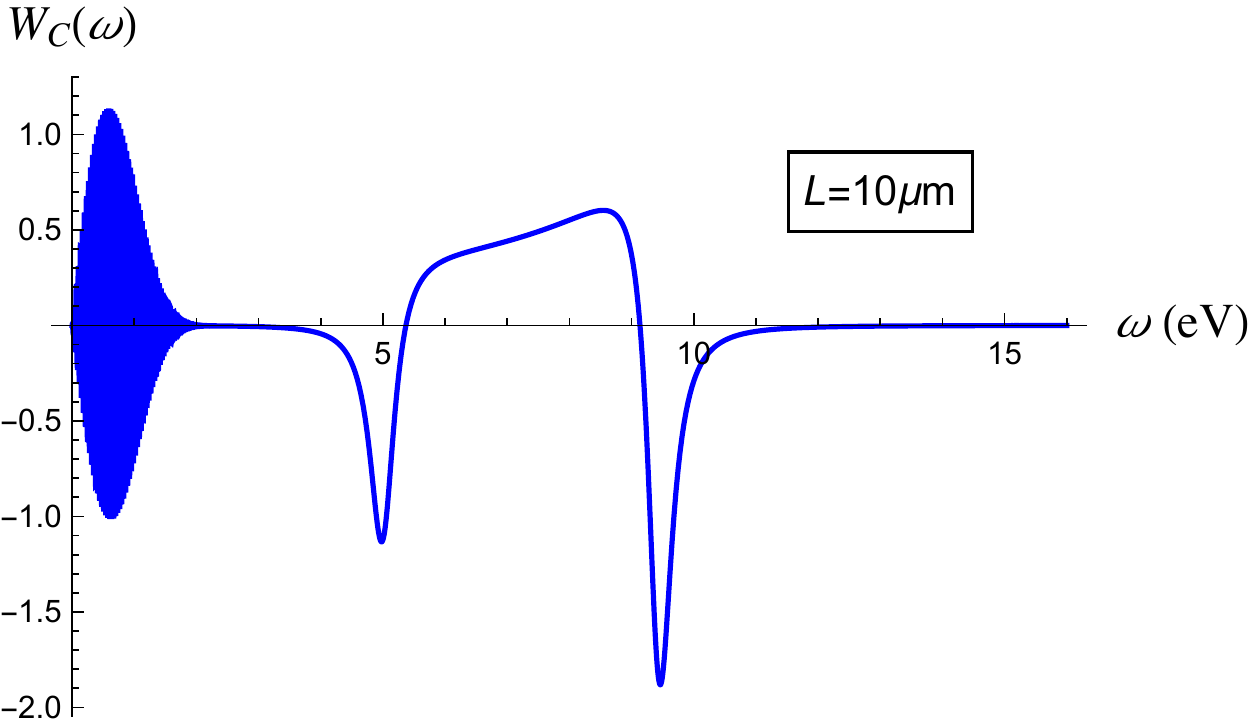}
\caption{Casimir energy per unit frequency (\ref{WC}) of the dielectric blocks described in Fig.~\ref{fig:diel1}. The total Casimir energy (\ref{encas}) is positive for all lengths $L$ of the dielectric.}
\label{fig:diel2}
\end{center}
\end{figure}

\section{Examples}
We now investigate the spectral energies $W(\omega)$ and $W_\mathrm{C}(\omega)$ for the cases where the block is a metal or a non-metallic dielectric. In both cases we choose $\mu(\omega)=1$ and a permittivity of the form
\begin{equation}  \label{osc}
\varepsilon(\omega)=1-\frac{\Omega^2}{\omega^2-\omega_0^2+\rmi\gamma\omega} .
\end{equation}
We put $\hbar=c=1$, with frequency in eV, length in $\mathrm{eV}^{-1}$, and the spectral energies $W(\omega)$ and $W_\mathrm{C}(\omega)$ dimensionless. 

For a metal we use a Drude-model approximation for gold~\cite{olm12}, namely (\ref{osc}) with $\omega_0=0$, $\Omega=8.45\,\mathrm{eV}$ and $\gamma=0.047\,\mathrm{eV}$. Figure~\ref{fig:gold1} shows the zero-point energy per unit frequency $W(\omega)$ contained in the block, for block lengths $L=1\,\mu\mathrm{m}$ ($5.068\,\mathrm{eV}^{-1}$) and $L=10\,\mu\mathrm{m}$ ($50.68\,\mathrm{eV}^{-1}$), together with the free-space value $\hbar \omega L/(2\pi c)$ of $W(\omega)$ (the spectral energy in the same spatial region without the block). We see that the block lowers the zero-point energies for frequencies less than the plasma frequency ($\omega\lesssim\Omega$), compared to the free-space values. (Recall that spatial regions outside the block do not contribute to changing the zero-point energy of the modes.) Above the plasma frequency the zero-point energies are larger than the free-space values. The Casimir energy per unit frequency $W_C(\omega)$ of the block is shown in Fig.~\ref{fig:gold2}, for block lengths $L=1\,\mu\mathrm{m}$ and $L=10\,\mu\mathrm{m}$. Above the plasma frequency the Casimir energy of the modes oscillates between positive and negative values, these oscillations in frequency becoming more rapid as $L$ increases. The free-space value of $W_C(\omega)$ is of course zero, so negative values of $W_C(\omega)$ correspond to a decrease of the regularized zero-point energy below that of empty space. The total Casimir energy (\ref{encas}) turns out to be positive for all $L$. 

For a non-metallic dielectric we use (\ref{osc}) with $\omega_0=5\,\mathrm{eV}$, $\Omega=8\,\mathrm{eV}$ and $\gamma=0.5\,\mathrm{eV}$ (these are not chosen to model a specific dielectric). Figures~\ref{fig:diel1} and~\ref{fig:diel2} show results for this dielectric that correspond to Figs.~\ref{fig:gold1} and~\ref{fig:gold2} for the metal. From Fig.~\ref{fig:diel1} we see that the zero-point energy of a range of frequencies is damped below the free-space value. The real part of the permittivity goes through zero at $\omega\approx5\,\mathrm{eV}$ and $\omega\approx10\,\mathrm{eV}$, and the damping of the zero-point energy occurs between these two frequencies. For parameters such that the real part of the permittivity (\ref{osc}) is positive for all frequencies, the decrease of zero-point energy below the free-space value occurs for frequencies around the resonance $\omega\approx\omega_0$. Figure~\ref{fig:diel2} shows that the Casimir energy per unit frequency $W_C(\omega)$ of the block oscillates between positive and negative values. As in the case of the metal, the total Casimir energy (\ref{encas}) is positive for all $L$. 

The effect of magnetic permeability and of negative refraction can also be investigated. Again, it is found that zero-point energy is suppressed below the free-space value for some frequencies and increased for others, while the Casimir energy per unit frequency oscillates between positive and negative values with the total Casimir energy being positive.

\section{Conclusions}
In quantum optics the electromagnetic vacuum (zero-point) field uncertainty of a subset of modes is given an experimental meaning through balanced homodyne detection~\cite{loudon} and spontaneous emission~\cite{pur46}. In such considerations the issue of regularization of the total (diverging) zero-point field uncertainty can be avoided, since the zero-point quantities for a single mode, or for a single frequency outside materials, are finite. Here we considered quantum optics in the presence of a block of material, taking full account of dispersion and absorption. We showed that the material decreases the zero-point energy of some modes, while increasing that of others. When the zero-point modes are regularized (in the same manner as is used to predict the Casimir force between real materials), then certain modes have negative Casmir energy while others have positive. The total Casimir energy is positive. 

The physics considered here is closely connected to that of nanomechanical systems, because it is nothing more than  quantum damped oscillators. Nanomechanical systems can be modelled using a Hamiltonian that differs from that of macroscopic electromagnetism in having one or more oscillators coupled to the reservoir instead of the infinite number of oscillators of the electromagnetic field~\cite{phi12,bar15}. In the simplest case of a single quantum damped oscillator similar modifications of zero-point energy can occur~\cite{phi12}. 

\section*{Acknowledgements}
I thank S.A.R. Horsley for helpful discussions.


\end{document}